# An air-spaced virtually imaged phased array with 94 MHz resolution for precision spectroscopy

Ibrahim Sadiek,[1,*] Norbert Lang,[1] and Jean-Pierre H. van Helden[1,2]

[1]*Leibniz Institute for Plasma Science and Technology (INP), 17489 Greifswald, Germany*
[2]*Experimental Physics V – Spectroscopy of Atoms and Molecules by Laser Methods, Faculty of Physics and Astronomy, Ruhr University Bochum, 44780 Bochum, Germany*
*\*ibrahim.sadiek@inp-greifswald.de*

**Abstract:** We report on an air-spaced virtually imaged phased array (VIPA) spectrometer that resolves the modes of a mid-infrared frequency comb with a repetition rate of 250 MHz, without an optical filter cavity. With a record spectral resolution of 94 MHz for VIPA, the spectrometer enables precision molecular spectroscopy with high resolution, broad spectral coverage and fast data acquisition. We demonstrate the capabilities of the spectrometer by measuring the absorption spectra of molecular species generated in plasmas. Using plasmas of a mixture of nitrogen, hydrogen, and methane at a low pressure of 1.5 mbar, we obtained high-resolution spectra of methane, around 3017 cm$^{-1}$, as well as hydrogen cyanide and ammonia, around 3240 cm$^{-1}$, demonstrating a wide spectral coverage over 290 cm$^{-1}$ (equivalent to 8.7 THz). The spectrometer performance in terms of Allan-Werle deviation and noise-equivalent absorption is discussed. The air-spaced VIPA concept offers a compact and practical spectrometer that harnesses the full potential of a stabilized frequency comb, making it suitable for a wide range of high-precision spectroscopic applications.

## 1. Introduction

A virtually imaged phased array (VIPA) is an innovative optical component that significantly enhances spectral resolution beyond traditional diffraction gratings [1]. A VIPA utilizes an etalon-based structure to create a phase array effect through multiple internal reflections. When coupled with a perpendicularly-aligned diffraction grating, unprecedented enhancement in the spectrometer resolving power is achieved [2]. This makes VIPAs essential in several applications, including efficient wavelength-division multiplexing in telecommunications [3], detailed imaging in optical coherence tomography [4], two-dimensional (2D) non-mechanical beam-steering for light detection and ranging (LiDAR) applications [5], and arbitrary optical waveform synthesis [6]. Their ability to enhance broadband absorption spectroscopy further underlines their versatility [2].

Traditionally, broadband spectroscopy relied on mechanical Fourier transform spectrometers (FTS) and dispersion-type grating-based spectrometers. Conventional FTS spectrometers, based on incoherent light sources, limit the spectral resolution, frequency accuracy and sensitivity of measurements [7]. Grating-based methods are usually faster and more compact than FTS; however, they provide poorer spectral resolution, typically on the order of a few GHz to tens of GHz [8].

The development of frequency combs as coherent broadband light sources has significantly improved the performance of these traditional broadband spectroscopic techniques in terms of spectral resolution, frequency accuracy and sensitivity [8–11]. The combination of frequency combs with mechanical FTS significantly reduced acquisition times by orders of magnitudes [12] compared to conventional FTS using incoherent light sources. By precisely matching the nominal resolution of the FTS to the repetition rate of the comb, the nominal spectrometer resolution set by the scanning range of the optical path difference was surpassed [13–15]. Alternatively, when two combs of slightly different repetition rates heterodyne against each other on a photodiode – dual comb spectroscopy [16–18] – a new comb in the RF domain enables broadband spectroscopic measurements with no mechanical moving parts at the video-rate speed [19].

Dispersive VIPA detection in frequency comb spectrometers is particularly attractive for recording spectra in the microsecond time scale, limited only by the integration time of the camera, and without involving mechanical moving parts. This makes VIPA-based systems ideal for high-resolution, fast data acquisition applications. Comb-based VIPA spectrometers have been used to investigate important molecular systems in atmospheric sciences like the Criegee intermediate [20] and dibromomethane [21], as well as for fundamental spectroscopy such as buckyball ($C_{60}$) [22]. Furthermore, they have been applied to rapid trace gas detection [23, 24], human breath analysis [25], and chemical kinetics [26, 27].

Previous implementations of comb-based VIPA spectrometers were constrained by limitations in resolving individual comb modes – due to either the necessity of using filtering cavities, which increased complexity, or due to the inherent resolution limits of the VIPA etalons, which reduced accuracy. Cavity-enhanced approaches, where cavities were used as optical filters, could resolve individual comb modes by increasing their spacing relative to the VIPA resolution [2, 28], and enabled spectra measurement with Hz-level precision [29]. However, the increased time needed to measure spectra at different repetition rates to resolve narrow absorption profiles makes the system more susceptible to long-term fluctuations. In contrast, cavity-free or cavity-enhanced implementations that lacked the ability to resolve individual comb modes have maintained the rapid acquisition times for trace gas detection [23, 24] and human breath analysis [25], as well as chemical kinetics [26, 27]. However, the spectral accuracy is reduced, particularly when the width of the measured absorption profiles is narrower than the intrinsic resolution of the VIPA system because of additional instrumental broadening. In such cases, a rigorous analysis is typically required to accurately evaluate the 'correct' instrument line shape function [30].

Here, we report on an air-spaced VIPA spectrometer that directly resolves the modes of low repetition rate frequency combs without the use of an optical filter cavity. With an effective resolution of $94 \pm 5$ MHz in the mid-infrared (mid-IR), this is to our knowledge the truly direct (cavity-free) comb-resolved measurements by a VIPA spectrometer. Air-spaced VIPA etalons outperforms traditional solid VIPA etalons, which suffer from significant temperature-induced tuning due to their large thermal expansion and thermo-optic coefficients. Previous work with solid VIPA etalons made of calcium fluoride [31] achieved an effective spectral resolution of approximately 490 MHz at 3 µm. This was the result of convolving the VIPA's intrinsic resolution of 190 MHz with an additional 450 MHz of instrumental broadening, which only partially resolved the modes of a 250 MHz repetition rate comb. Another approach in cross-dispersive spectroscopy used an immersion grating instead of the conventional solid VIPA etalons and achieved a similar spectral resolution of 460 MHz at 9 µm [32]. Our air-spaced VIPA spectrometer offers 8.7 THz of spectral coverage combined with rapid data acquisition, enabling detailed analysis of multiple molecular species and overcoming the limitations of slower, traditional broadband spectroscopic methods. In addition, it achieves broad spectral coverage, high spectral resolution, and fast data acquisition in a compact setup. This ensures that precision spectroscopy can be achieved in both cavity-free and cavity-enhanced forms across a variety of applications, including in challenging molecular environments.

While most applications of direct frequency comb spectroscopy for molecular absorption measurements in the literature were performed under thermal conditions [8–12, 33–37], there have been a few instances where frequency comb spectroscopy was applied to study highly non-thermal molecular plasmas [38–41]. Molecular plasmas are utilized here as a model system for complex chemical processes of industrial importance [42], and to create molecular species that exhibit absorptions over a broad spectral range. We demonstrate the high spectral resolution, broad spectral coverage, and fast data acquisition achievable with the compact air-spaced VIPA by measuring the absorption spectra of a plasma containing a mixture of nitrogen ($N_2$), hydrogen ($H_2$), and methane ($CH_4$) under low pressure conditions. Absorption spectra of plasma-generated species like $CH_4$ around 3017 cm$^{-1}$ and hydrogen cyanide (HCN) and ammonia ($NH_3$) around 3240 cm$^{-1}$, were measured precisely, demonstrating the broad spectral coverage of the spectrometer.

## 2. Experimental details

The overall experimental setup is shown in Fig. 1(a) and it consists of a mid-IR frequency comb, a plasma reactor, and the air-spaced VIPA detection system. The mid-IR comb (Menlo Systems), emitting in the 2800 – 3400 cm$^{-1}$ range and capable of exciting the C-H and N-H vibrations of molecular species produced in the plasma. The mid-IR radiation was generated through difference frequency generation in a periodically poled lithium niobate crystal. This was achieved by combining a spectrally shifted copy at 1040 nm of a high-power Er-fiber comb around 1550 nm with the original comb at a repetition rate, $f_{rep}$, of 250 MHz. Consequently, the comb was inherently free from the carrier envelope offset frequency. The $f_{rep}$ of the comb was locked to the output of a tunable direct digital synthesizer referenced to a GPS-disciplined quartz oscillator. A pair of single mode ZBLAN ($ZrF_4$, $BaF_2$, $LaF_3$, $AlF_3$ and NaF – Thorlabs) fibers and two pairs of fiber couplers were used to couple the comb light into the reactor and out of it into the detection unit (only one is shown in the schematic of Fig. 1(a)). The laser beam passed twice through the reactor via a fixed retroreflector at the back of the reactor resulting in an effective absorption path length of 163.5 ± 0.7 cm. Another motorized retroreflector was used to bypass the plasma reactor for reference measurements.

The plasma was generated in a DC discharge, operated with a mixture of $N_2$, $H_2$, and $CH_4$ gases at a low pressure of 1.5 mbar. The pressure was maintained using a back pressure controller and vacuum pump. The reactor was a specially designed and scaled-down version of an industrial scale reactor used for hardening stainless-steel and tools by nitrocarburizing processes [43]. Details of the reactor can be found elsewhere [40]. After passing through the plasma reactor, the frequency comb beam was coupled to the compact setup of the air-spaced VIPA detection unit mounted on an optical breadboard of 70 cm × 40 cm. The main components of the VIPA detection system include a cylindrical lens, a VIPA etalon, a grating, and an IR camera. We used a customized air-spaced VIPA etalon (LightMachinary) constructed of two mid-IR transparent material windows with a coating optimized for 3.2 μm (or 3125 $cm^{-1}$). A schematic of the air-spaced VIPA etalon is shown in Fig. 1(b). The two optical windows, each with a clear aperture of 40 mm × 80 mm, were separated by 37.5 mm and mounted together with Zerodur spacers resulting in a free spectral range (FSR) of 4 GHz. In the VIPA spectrometer, the comb beam, after being line-focused by the cylindrical lens (focal length = 7.5 cm), propagates between the two VIPA windows and therefore is dispersed vertically with the different wavelengths being dispersed at different angles.

Measurements of the propagated comb beam within the VIPA detection unit using an IR beam profiler (DataRay Inc.) are shown Fig. 1(c) at different locations: (I) before the cylindrical lens; (II) within approximately 1 cm of the focal point of the cylindrical lens; and (III) right after the air-spaced VIPA etalon. The last profile was measured at a relatively large tilt angle of the VIPA etalon of approximately 2.5°. In order to disperse the frequency comb beam vertically as implemented in the end in the VIPA spectrometer, the tilt angle was decreased to approximately 0.5° and then the profile of (III) formed a vertical stripe. The vertical stripe was subsequently cross-dispersed via a 100 mm wide Echelle grating (Richardson Gratings) with 112 lines/mm at its $5^{th}$ order. The resulting 2D dispersion images were recorded with a mid-IR camera (InSb, InfraTec) with a dynamic range of up to 16 bits, an array of (512 × 640) IR-pixel, a spectral sensitivity range of 1.5 – 5.7 μm, a pitch size of 20 μm, and a minimum integration time of 500 ns. Figure 1(d) shows a typical 2D dispersion image recorded with the air-spaced VIPA spectrometer for a frequency comb with an $f_{rep}$ of 250 MHz. It clearly demonstrates that the spectrometer is capable of resolving the comb modes without having an optical filter cavity in the setup; the 'dots' are separated vertically by 250 MHz and horizontally by 4000 MHz.

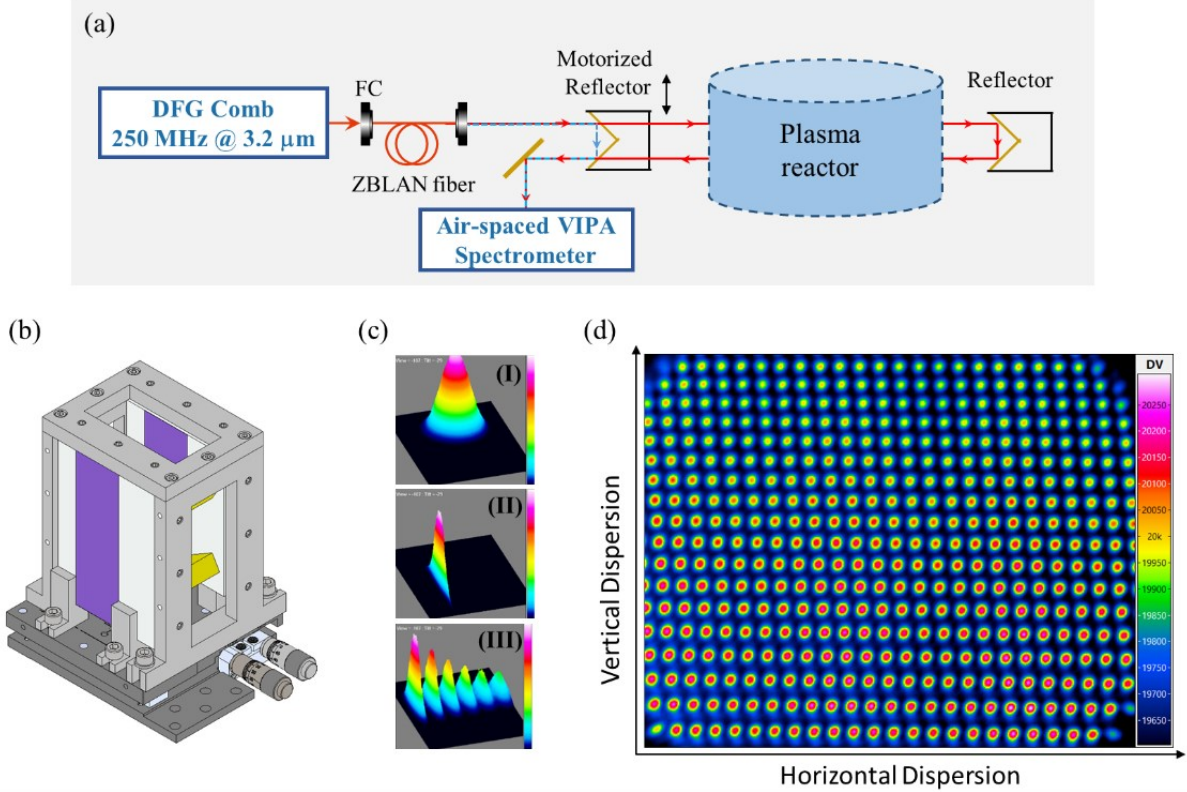

**Fig. 1.** (a) Schematic of the overall experimental setup, consisting of a mid-IR frequency comb, a plasma reactor, and an air-spaced VIPA detection unit. A pair of ZBLAN fibers and two pairs of fiber couplers (FC) were used to couple the comb to the reactor and from the reactor to the detection unit. (b) A perspective view of the customized air-spaced VIPA etalon in an optomechanical housing (the reflective coatings are shown in purple and the spacers in yellow). (c) Beam profiles measured at different places in the VIPA setup: (I) before the cylindrical lens, (II) close to the focal length of the cylindrical lens, and (III) after the air-spaced VIPA etalon at atilt angle of the VIPA etalon of ∼ 2.5°. (d) A measured 2D image with an IR camera, showing the resolved modes of a mid-IR comb with an $f_{rep}$ of 250 MHz.

The air-spaced VIPA spectrometer allows the simultaneous acquisition of spectral ranges of up to 8 cm$^{-1}$, corresponding to about 960 spectral elements within one camera frame. The grating was mounted on a motorized rotational stage, enabling a sequential collection of up to 50 cm$^{-1}$ (∼ 6000 spectral elements) without adjusting the alignment. The mount of the grating allows for roll and pitch rotation as well as for translation, which was useful to avoid large tilts in the fringes (or stripes) that can cause steps in the baseline of the measured spectra. Similar effects were observed in [24], and it was compensated differently by rotating the camera in [44].

At each grating angle, three types of frames were acquired: a signal frame ($I_{sig}$), with the plasma switched on and the laser beam passing through the reactor, a reference frame ($I_{ref}$), with the laser beam bypassing the reactor, and a camera offset frame ($I_{off}$), with 500 frames averaged per measurement. With an integration time of 478 µs and an acquisition rate of 40 Hz (equivalent to 25 ms), the measurement time per spectral window was 37.5 s. The intensities of the pixels were sampled at the comb mode positions within one VIPA FSR following the approach outlined in Ref.[29]. In that approach, a grid along the dispersion pattern from unresolved comb images (i.e., without using filtering cavity) was created. This grid then served as a guide along which the subsequently measured images containing resolved comb modes obtained by optical cavity filtering were analyzed. In our work, all the measured images are for resolved comb modes. Therefore, we created the grid along the vertical stripes by interleaving of four images of resolved comb modes measured at different $f_{rep}$. Within this grid, the maxima of the pixel intensities in the resolved comb mode images from single $f_{rep}$ measurements were located and the comb teeth intensities were integrated. For each comb mode the integral of 21 pixels was used to obtain the pixel intensity. The absorption spectrum was then calculated by normalizing the signal spectrum, $I(\tilde{\nu}) = I_{sig}(\tilde{\nu}) - I_{off}(\tilde{\nu})$ by a reference spectrum $I_0(\tilde{\nu}) = I_{ref}(\tilde{\nu}) - I_{off}(\tilde{\nu})$. Baseline correction was applied using a third order polynomial and a number of sine terms to remove baseline etalons.

## 3. Results and discussion

In contrast to previous VIPA implementations that utilized solid etalons, we employed a customized air-spaced VIPA to resolve the modes of a low repetition rate comb of 250 MHz. We applied this spectrometer for precision spectroscopy in molecular plasmas in the mid-IR region, around $\tilde{\nu}$ = 3125 cm$^{-1}$. Figure 2(a) shows a measured 2D dispersion image at one $f_{rep}$ of the comb of 250 MHz. The spectrum was measured at a plasma power of 68 ± 1 W with gas flow rates of 20 standard cubic centimeters per minute (sccm) $N_2$, 20 sccm $H_2$, and 2 sccm $CH_4$ at a total pressure of 1.5 mbar.

As shown in this figure, the intensities of some of the resolved comb modes are reduced due to the absorption by $CH_4$ transitions, around 3017 cm$^{-1}$, where the Q-branch of the $\nu_3$ band of $CH_4$ exhibits strong absorptions. Note that $CH_4$ was added as a precursor gas, and therefore it will get dissociated, creating other carbon-containing molecules (e.g., $C_2H_2$, $C_2H_4$, and $C_2H_6$), in addition to the possible reformation of $CH_4$. Additionally, other nitrogen-containing molecules such as HCN and $NH_3$ are also generated in the plasma (*see below*). The distance between the two dashed lines in Fig. 2(a) marks one FSR of the VIPA etalon and it covers approximately 57 % of the detector area. With the FSR spanning about 290 pixels, a frequency per pixel of approximately 13.8 MHz can be achieved, which ensures that no instrumental broadening by the camera array when measuring resolved comb modes with the air-spaced VIPA of a spectral resolution of 94 ± 5 MHz. The resolution of the VIPA is determined from the full width at half maximum (FWHM) of Gaussian fits to the resolved comb modes along one stripe in the center of the image. It is worth noting that a Gaussian function provides a better fit to the resolved comb modes in Fig. 2(a) than a Lorentzian function. We also note that full 2D images taken by the camera array showed an almost circular crop, similar to what has been measured elsewhere [30]. We could later reduce this crop by the addition of an achromatic doublet lens in the imaging arm in front of the camera.

For precise absorption profile measurements, we interleaved a total of four spectra measured at different $f_{rep}$ separated by 173.75 Hz in the frequency domain, corresponding to a point spacing of 62 MHz in the optical domain. Figure 2b shows the final interleaved spectrum of $CH_4$ around 3017 cm$^{-1}$. The two spectral windows, blue and green, correspond to measurements at two different angles of the grating. As the comb modes are resolved, a comparison of the analyzed images (which should contain at least one known absorption line) with spectra from the HITRAN2020 database [45] allows for the sequential integer numbers $n$ to be determined for each frequency comb tooth in the image. Therefore, the center frequency of each comb tooth has an accuracy limited by the reference clock used for the $f_{rep}$ stabilization. Here, the absolute frequency axis was calibrated to the absorption lines of $CH_4$ and HCN obtained from the HITRAN2020 database [45].

Also shown in Fig. 2(b) is a transmission model (red, inverted for clarity) based on the HITRAN2020 database and a Voigt line shape function, together with the fit residuals shown in the lower panel. In the spectral model, the line positions, the pressure line broadenings and shifts were fixed to their values from the HITRAN database [45], while the Doppler width was set as a free parameter to determine the mean kinetic energy of species in the plasma. The Doppler width of transitions was determined by broadband fit, i.e., by fitting a global scaling factor to their theoretical values of the Doppler widths calculated at the listed line centers from the HITRAN2020 database and at 294 K. We determined a Doppler width of 0.0097(2) cm$^{-1}$ for methane, which corresponds to a gas temperature of 316 ± 7 K. A zoomed-in spectral window in the overlapping region between the two datasets, green and blue, measured at different grating angles is shown in Fig. 2(c). We note that the spectrometer is capable of resolving lower wavenumbers beyond those indicated in Fig. 2(b), limited only by the laser bandwidth. To evaluate whether the measured profiles include additional instrumental broadening, we performed a separate measurement of methane at gas flow rates of 1 sccm of $CH_4$, 20 sccm of $N_2$ and 20 sccm of $H_2$, and at a total pressure of 1.5 mbar, but with the plasma switched off to avoid any additional inhomogeneous broadening by the plasma heating. An additional broadening of Lorentzian type was included in the model as a global fitting parameter, resulting in a FWHM of ~12 MHz (or 4 ×10$^{-4}$ cm$^{-1}$). This additional broadening is ~ 4 % of fitted Doppler width. The consideration of this broadening reduced the noise level, evaluated as the standard deviation of the residuals, by a factor of ~1.4. We attribute this additional broadening to cross-talk between resolved comb modes, where a mixing of 2 % between adjacent modes is predicted to introduce ~ 16 MHz broadening in the measured profiles. The residuals (Exp. – Fit) shown in Fig. 2 are for the fitting model including the 12 MHz cross-talk broadening. From two measurements, the decrease in the absorbed signal strength of methane when plasma was switched on can be related to a depletion of up to 60 % of methane.

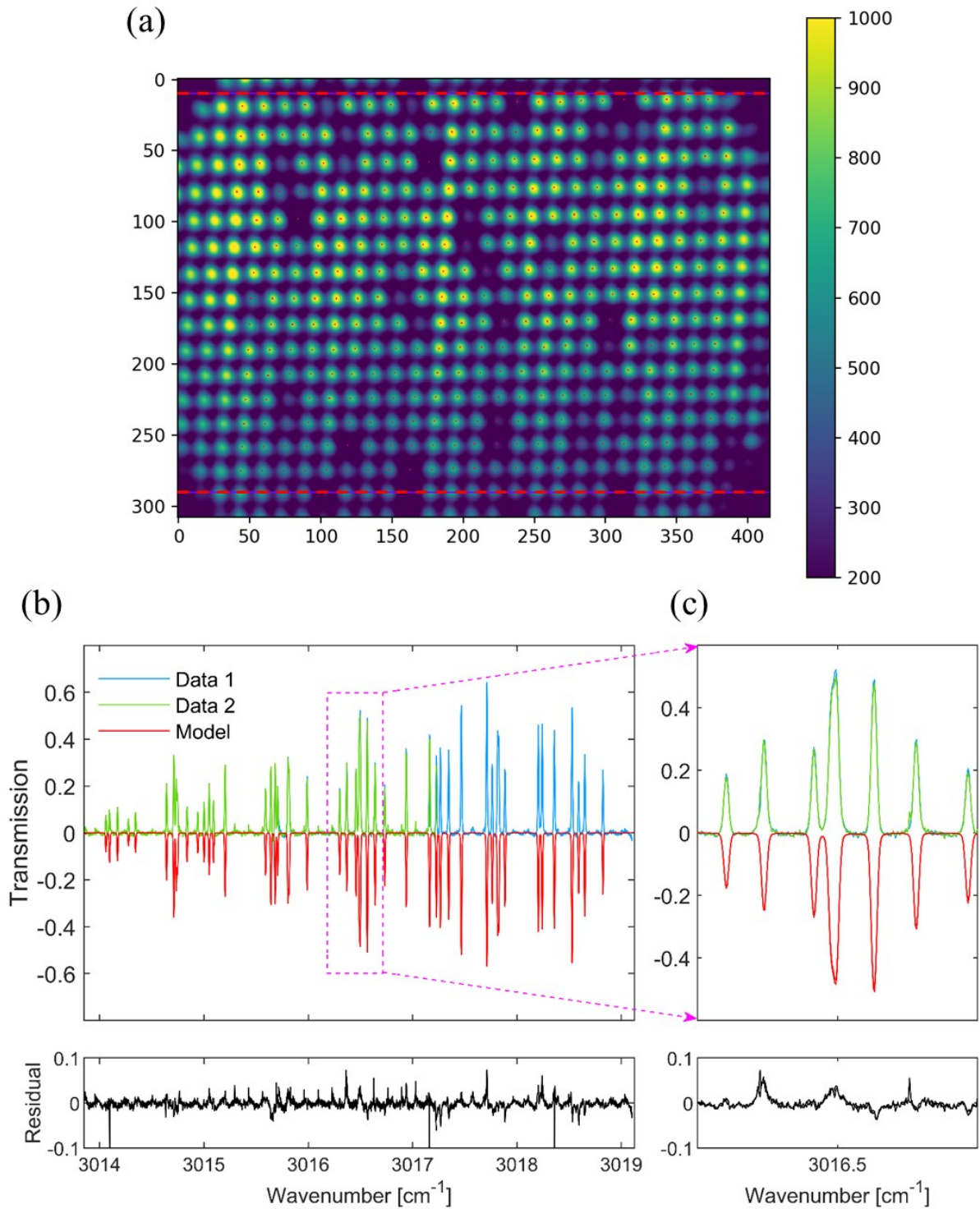

**Fig. 2.** (a) 2D signal image (500 averages, 40 Hz collection rate) collected with the air-spaced VIPA spectrometer, showing resolved modes of the frequency comb ($f_{rep}$ = 250 MHz). The horizontal dashed lines indicate the VIPA FSR. The frames were collected while the plasma was switched on with mass flow rates of 20 sccm $N_2$, 20 sccm $H_2$, and 2 sccm $CH_4$, at a total pressure of 1.5 mbar, and a plasma power of 68 ± 1 W. (b) High-resolution absorption spectrum obtained from measurements at four different repetition rates. At each $f_{rep}$, signal, reference and background frames

were measured to create the final spectrum. Data 1 (blue) and Data 2 (green) correspond to measurements at different grating angles. Also shown in (b) is a fitting model for $CH_4$ (red, inverted for clarity) based on the HITRAN2020 database [45] using a Voigt line shape function. (c) Zoomed-in spectral window at the overlapping region of Data 1 and Data 2. The fit residuals are shown in the lower panels.

To demonstrate the wide spectral coverage of the spectrometer, we measured the rovibrational transitions of the $\nu_1$ band of reactive $H^{12}C^{14}N$ generated in the plasma around 3290 cm$^{-1}$. Figure 3(a) shows the measured transmission spectra (black) for two grating angles. The rovibrational absorption features of HCN are about 290 cm$^{-1}$ apart from the spectra of $CH_4$ presented in Fig. 2, indicating the broad spectral coverage achieved by the air-spaced VIPA spectrometer. As highlighted earlier, the spectrometer's spectral coverage is limited by the laser bandwidth, indicating that the spectral range could be extended to higher wavenumbers, up to 3400 cm$^{-1}$. Here, it is demonstrated up to 3300 cm$^{-1}$. Also shown in Fig. 3(a) are the fit residuals (Exp.– Fit, yellow) to a model based on the HITRAN database [45] and a Voigt line shape function. The pressure line broadenings and shifts were fixed to their values from the HITAN2020 database [45], and including an instrumental broadening of 12 MHz as discussed above. The Doppler width was fitted as described earlier. From the broadband fit a translational temperature of 476 ± 4 K is determined for HCN in the plasma. This value also agrees with the value of 474 ± 7 K determined from Doppler widths obtained from the fitting of Voigt functions to individual transitions presented in Fig. 3(a). Figure 3(b)–(d) show zoomed-in spectral windows with resolved absorption profiles of $H^{12}C^{14}N$, Fig. 3(b) and Fig. 3(d), and $^{14}NH_3$, Fig. 3(c), together with the model (red) and the fit residuals (E – F, yellow).

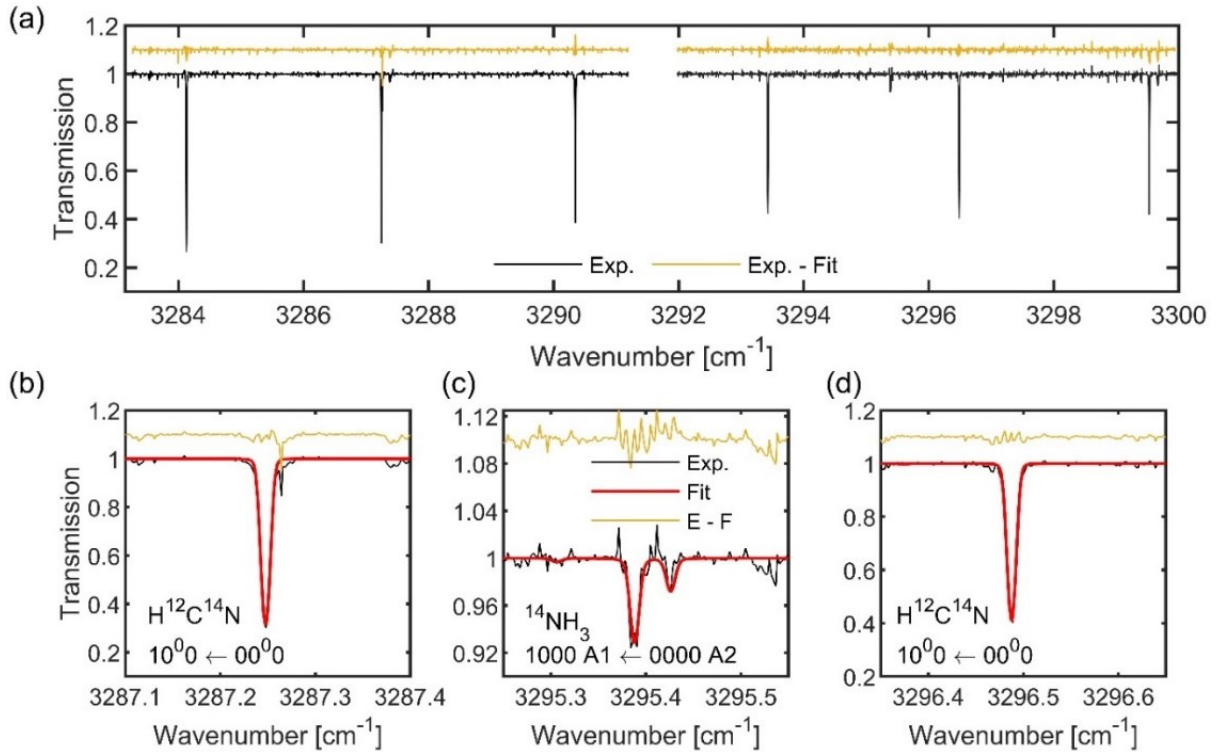

**Fig. 3.** (a) Measured (black) high-resolution spectrum of $H^{12}C^{14}N$ generated in the plasma (gas flow rates: 20 sccm $N_2$, 20 sccm $H_2$, and 2 sccm $CH_4$; total pressure: 1.5 mbar; plasma power: 68 ± 1 W). The two spectral regions around 3292 cm$^{-1}$ correspond to measurements at two grating angles. Also shown in (a) is the fit residual (Exp. – Fit, yellow) from a model based on the HITRAN2020 database [45] and a Voigt line shape function. (b) – (d) Zoomed-in spectral windows with resolved rovibrational profiles for $H^{12}C^{14}N$ (b & d) and $^{14}NH_3$ (c).

To evaluate the noise in the spectrometer, we calculated the Allan-Werle deviation of a single spectral element from frames measured at a data collection rate of 40 Hz. Fig. 4 presents the results of the Allan-Werle deviation plot in terms of absorbance = $\alpha \times L$, where $\alpha$ is the absorption coefficient and $L$ is the pathlength. As can be seen from the comparison with the theoretical characteristic curve for white-noise (red line with a slope of – 0.5), the noise decreases statistically up to one second measurement time, which demonstrates the very good stability of the spectrometer. The deviation after one second is attributed to slow drifts in the system, including drifts in the camera dark currents and in the baseline etalons. This optimum time of one second corresponds to an absorbance of $1.6 \times 10^{-3}$. This value compares very well with those reported in Table 5 of Ref.[46], for a variety of comb-based methods which are in the range of $8.6 \times 10^{-5} – 4.1 \times 10^{-2}$ for a one second measurement time. Among this list [46], a minimum absorbance of $6.0 \times 10^{-3}$ was reported for a high-sensitivity cavity-enhanced direct frequency comb experiment with a VIPA detection system. This is about a factor of 3 higher than the achieved minimum absorbance in this study. In terms of the minimum absorption coefficient, $\alpha_{min}$, the 1 s measurements time corresponds to a value of $9.8 \times 10^{-6}$ cm$^{-1}$. This yields a noise equivalent absorption sensitivity, $NEAS = \alpha_{min} \times t^{1/2}$, of $6 \times 10^{-5}$ cm$^{-1}$ Hz$^{-1/2}$, where the measurement time $t$ was 37.5 s for obtaining a spectrum at one grating angle.

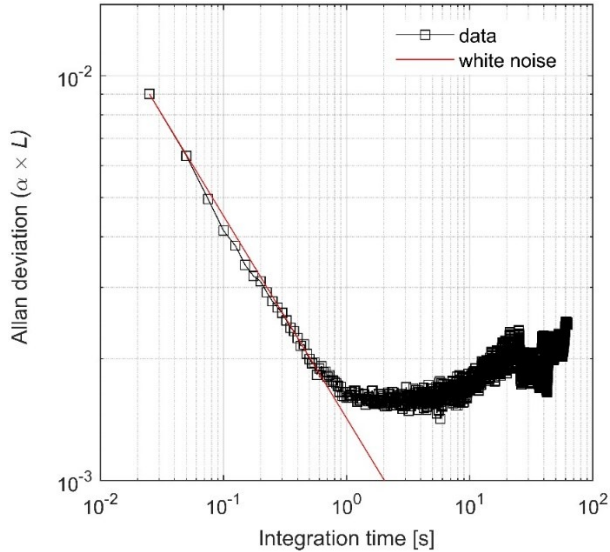

**Fig. 4.** Allan-Werle deviation (black) in terms of absorbance = $\alpha \times L$, for a single spectral element of resolved comb mode in the 3017 cm$^{-1}$ spectral region, where the comb has the highest intensity. Also shown is a red line with a slope of −0.5, representing the characteristic of white noise.

## 4. Conclusions

We report on an air-spaced Virtually Imaged Phased Array (VIPA) spectrometer with a record spectral resolution of 94 ± 5 MHz, which to the best of our knowledge, is the highest reported resolution of a VIPA spectrometer for precision spectroscopy. This spectrometer can resolve modes of low repetition rate combs directly, without the need for filtering cavities. Here, we demonstrate this capability by resolving the modes of a mid-IR comb with a repetition rate of 250 MHz, with minimal instrumental broadening of less than 4 % of the measured absorption profiles.

Previous work with solid VIPA etalons achieved an effective spectral resolution of approximately 490 MHz at 3 µm, despite an intrinsic resolution of 190 MHz, which only partially resolved the modes of 250 MHz repetition rate comb. The air-spaced VIPA spectrometer, by contrast, resolves the modes of such low-repetition-rate combs without significant instrumental broadening. It outperforms solid VIPA etalons, which are affected by high etalon-temperature tuning rates due to large thermal expansion and thermo-optic coefficients. Such large thermal expansion of solid VIPA etalons, particularly at such high spectral resolution, can impair the spectrometer's performance, as changes in resolution due to temperature fluctuations are more pronounced relative to the etalon's intrinsic resolution. Furthermore, the air-spaced VIPA spectrometer enables precision spectroscopy in both cavity-free implementations (which previously could not resolve comb modes) and cavity-enhanced setups (which required filtering cavities for comb mode resolution). We demonstrated these capabilities by measuring multiple molecular species over a broad spectral coverage of 8.7 THz in the mid-infrared in complex molecular environments of plasmas.

**Funding.** Deutsche Forschungsgemeinschaft (DFG, German Research Foundation) – project No. SA 4483/1–1. German Federal ministry of Education and Research – project No. 13N14947.

**Acknowledgement.** The authors acknowledge the technical support and valuable comments from G. Kowzan and P. Masłowski, and the valuable comments from A. J. Fleisher. Additionally, we thank U. Macherius and F. Weichbrodt for the technical support.

**Author contributions.** According to CRediT (Contributor Roles Taxonomy), **I. Sadiek.** has contributed to Conceptualization, Formal analysis; Investigation; Visualization; Methodology and Writing – original draft, **N. Lang** has contributed to Project administration; Data curation; Methodology; Conceptualization and Writing – review & editing, **J. H. van Helden** has contributed to Resources; Methodology; Conceptualization and Writing – review & editing.

**Disclosures.** The authors declare no conflicts of interests.

**Data availability.** All data that support the findings of this study are available within the paper and also available from the corresponding author upon reasonable request.